\documentclass[twocolumn,showpacs,amsmath,amssymb,superscriptaddress]{revtex4-1}

\usepackage{graphicx} 
\usepackage{dcolumn} 
\usepackage{bm} 
\usepackage[caption=false]{subfig}
\usepackage{color}

\newcommand{\refb}[1]{Fig.~\ref{#1}} 

\begin{document}

\title{Jammed frictional tetrahedra are hyperstatic}

\author{Max Neudecker}
\email{max.neudecker@ds.mpg.de}
\affiliation{Max Planck Institute for Dynamics and Self-Organization (MPIDS), 37077 Goettingen, Germany}
\author{Stephan Ulrich}
\email{Present address: Instituut-Lorentz, Leiden University}
\affiliation{University Goettingen, Institute of Theoretical Physics, Germany}
\author{Stephan Herminghaus}
\author{Matthias Schr\"oter}
\email{matthias.schroeter@ds.mpg.de}
\affiliation{Max Planck Institute for Dynamics and Self-Organization (MPIDS), 37077 Goettingen, Germany}

\date{\today}

\begin{abstract}
We prepare packings of frictional tetrahedra with volume fractions $\phi$ ranging from $0.469$ to $0.622$ 
using three different experimental protocols under isobaric conditions.  
Analysis via X-ray microtomography reveals that the contact number $Z$ grows with $\phi$, 
but does depend on the preparation protocol. While there exist four different types of contacts in tetrahedra packings, 
our analysis shows that the edge-to-face contacts contribute about 50\% of the total increase in $Z$.  
The number of constraints per particle $C$ increases also with $\phi$ and 
even the loosest packings are strongly hyperstatic i.e.~mechanically over-determined with $C$  
approximately twice the degrees of freedom each particle possesses.
\end{abstract}



\maketitle

\begin{figure}
  \begin{center}
  \subfloat{\label{fig:particles_photo}\includegraphics[width=27mm,height=27mm]{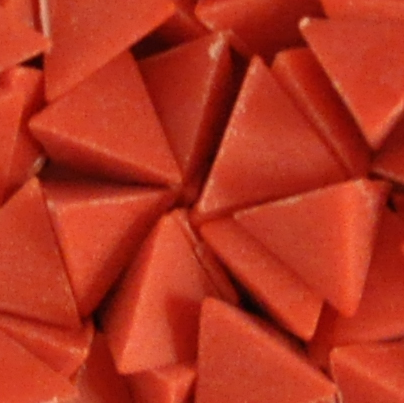}}
  \hspace{0.0cm}
  \subfloat{\label{fig:particles_rendered}\includegraphics[width=27mm,height=27mm]{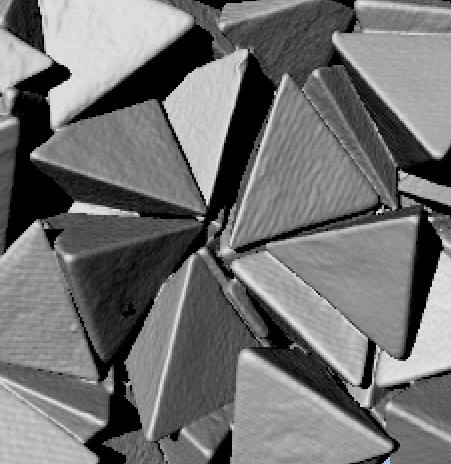}}
  \hspace{0.0cm}
  \subfloat{\label{fig:particles_povray}\includegraphics[width=27mm,height=27mm]{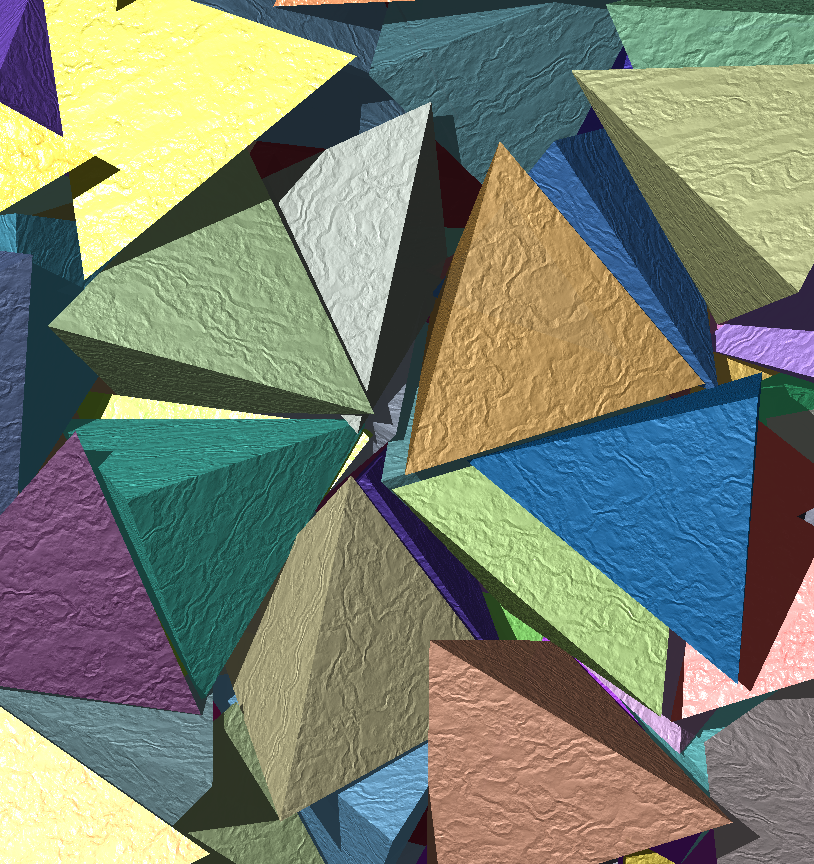}}
  \caption{From left to right: Close-up view of the surface of an experimental tetrahedra packing; 
    Rendering of X-Ray Computed-Tomography data; Representation by exact tetrahedra, colored by orientation. (Color online)
}
\label{fig:particles} 
 \end{center}
\end{figure}

{\it Introduction. --}
The jamming paradigm \cite{liu:10,vhecke:10} provides a unified viewpoint on 
the onset of mechanical stability in particulate soft matter systems such as wet foams, emulsions, and colloids.
All those systems are composed of frictionless spherical particles, consequentially they share
the existence of a critical volume fraction 
$\phi_J \approx 0.64$ where the system becomes isostatic, i.e.~the number of constraints
imposed by the contacts corresponds to the degrees of freedom of the particles. Higher densities
can only be achieved by increasing the pressure which leads to excess contacts;
these excess contacts then determine the mechanical properties of the system. 

Granular matter differs in two ways from these systems.
First, its nonzero surface friction $\mu$ allows for tangential forces at 
particle contacts. These result in mechanically stable packings over a range of $\phi$ 
{\it at any given pressure}. Such packings differ in their structure, not their compression.
However, one can still define a meaningful distance to isostaticity \cite{shundyak:07,henkes:10}. 

Second, most real world granular materials like sand, sugar or soil are not spherical. 
Only quite recently the jamming approach has been applied to frictionless ellipsoids \cite{zeravcic:09,mailman:09}
and tetrahedra \cite{smith:10}. Tetrahedra provide besides jamming a number of other interesting packing problems
like the existence of  quasi crystals \cite{haji-akbari:09} or the quest
for the densest packing \cite{torquato:09,kallus:10,torquato:10}, which presently is
believed to be a ``dimer crystal'' with $\phi \approx 0.8563$ \cite{chen:10}.
However, experimental packings of (necessarily) frictional tetrahedra explore a much lower range of $\phi$ \cite{baker:10}.

From a jamming point of view, tetrahedra introduce a new element: 
unlike spheres they have four different types of contacts: face to face, edge to face, edge to edge, and vertex to face.
Nevertheless, previous numerical results have found frictionless uncompressed packings of tetrahedra 
to be isostatic \cite{smith:10,smith:11,jiao:11}; the same result has been reported by the only 
experiment so far on this topic, where the
 contact numbers of frictional tetrahedral dice were determined\cite{jaoshvili:10}. 

In this letter, we use X-ray tomography to determine the particle positions and orientations inside tetrahedra packings. 
In contrast to  reference \cite{jaoshvili:10}, even our loosest packings are highly hyperstatic. We also show that 
the contact numbers are  protocol-dependent. 

{\it Experimental setup. --}
The tetrahedral particles in our study are produced by injection moulding of polypropylene, and have a side length of $a=7\,\mathrm{mm}$ 
(see Fig.~\ref{fig:particles} left). The typical radius of curvature of edges and corners is $150 \pm 50\,\mathrm{\mu m}$ which corresponds 
to 2.5\% of $a$ (for tetrahedral dice, as used  in \cite{jaoshvili:10,baker:10}, this ratio is 6-8\%).
All particles have small mold marks on one of the 6 edges, the height of this protrusions  
amounts to $ 300 \,\mathrm{\mu m}$ at maximum \cite{note1}. The particles have a coefficient of friction $\mu=0.87 \pm 0.03$,
which is measured using a tilted plane and finding the angle at which face-to-face contacts start to slide past each other
(tetrahedral dice: $\mu=0.22 \pm 0.04$).

In order to explore the structure of tetrahedral packings at different densities, we compactify them by vertical shaking.
We start from initially loose packings which are prepared by filling 
$14406$ particles via a slowly lifted  funnel into a cylinder of inner diameter
 $D=104\, \mathrm{mm}$ ($\approx 15\,a$).  
This method creates packings with $\phi$  between 0.469 and 0.470, the lowest values found in this study.  
An electromagnetic shaker (LDS V555) then applies sinusoidally shaped pulses (``taps'') with a width of $50\,\mathrm{ms}$ and a 
repetition rate of $3\,\mathrm{Hz}$. 
The peak acceleration $A_p$ per tap is measured at the base plate of the container (Kistler 8763B) and reported in 
units of the dimensionless acceleration $\Gamma = A_p/g$ (where $g=9.81 \mathrm{m/s^2}$).

A laser distance scanner (MicroEpsilon ILD1402) mounted on a translation stage measures the surface height profile after shaking; 
from this profile the corresponding $\phi$ is computed using calibration measurements from tomographic reconstruction.  
This results in an accuracy of the reported $\phi$ values of $\pm$ 0.003. 
Tomograms of the prepared packings are acquired using X-ray Computed Tomography  (GE Nanotom) with a tungsten target and  
$120\,\mathrm{kV}$ acceleration voltage. 
Each dataset comprises 1152$\times$1152$\times$1076 voxels with a resolution of $100\,\mathrm{\mu m}$/voxel.

\begin{figure}
\subfloat{\label{fig:compaction_loose_dense}\includegraphics[width=0.24\textwidth]{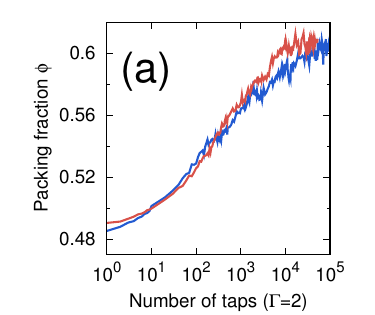}}
\hspace{-0.5cm}
\subfloat{\label{fig:compaction_taps}\includegraphics[width=0.24\textwidth]{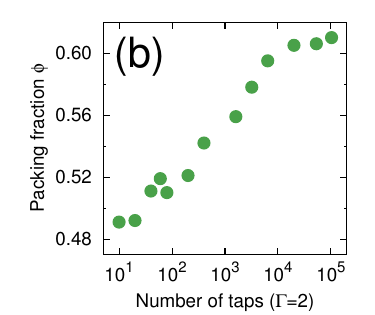}}  \\
\vspace{-0.0cm}
\hspace{-0.5cm}
  \subfloat{\label{fig:compaction_annealing}\includegraphics[width=0.48\textwidth]{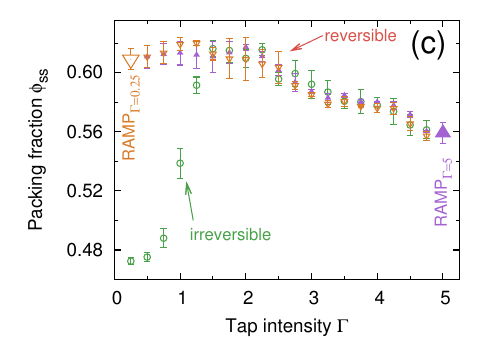}}
\caption{(a) Protocol TAP: Tapping at $\Gamma=2$ compacts the sample until it reaches a steady state
after approximately 10$^4$ taps (two independent runs are shown).
(b) By stopping after a given number of taps, we obtain 13 different samples which we analyze by tomography.
(c) Protocol RAMP: At each acceleration step, the packing is tapped $10^4$ times. 
Open circles (green): initial, irreversible compaction. Open downward triangles (orange): decreasing acceleration. 
Filled upward triangles (violet): increasing acceleration. 
Packings at points RAMP$_{\Gamma=0.25}$ and RAMP$_{\Gamma=5}$ are analyzed via tomography.
All error bars are averaged over 4 independent runs. (Color online)
}
\end{figure}

{\it Preparation protocols. --}
We use three different preparation protocols.
The first one will be referred to as TAP: 
Starting from a loose packing, a number of taps ranging from $10^1$ to $10^5$ 
with a constant acceleration $\Gamma=2$ is applied.
As \refb{fig:compaction_taps} and \refb{fig:compaction_loose_dense} show,
the packing fraction increases monotonely with the number of taps, 
reaching a steady state after approximately $10^4$ taps with a packing fraction $\phi_{\mathrm{ss}}$.

Our second protocol RAMP is intended to study the possible range of steady state volume fractions $\phi_{\mathrm{ss}}$.
Following Nowak \textit{et al.}\cite{nowak:98},  the tap intensity $\Gamma$ is increased stepwise from 0.25 to 5, decreased back to 0.25, 
and increased to 5 again.  At each step, $10^4$ taps are applied and the resulting  $\phi_{\mathrm{ss}}$ is measured. 
The evolution of $\phi_{\mathrm{ss}}$ is depicted in \refb{fig:compaction_annealing},
and comprises the following regimes: The loose packing irreversibly compacts when $\Gamma$ is increased from 0.25 to 1.5, but
 dilates again when the acceleration is further ramped to up $\Gamma=5$, leading to $\phi_{\mathrm{ss}}\approx 0.56$. 
Subsequently, decreasing the tap intensity down to $\Gamma=0.25$ creates a dense packing with $\phi_{\mathrm{ss}} \approx 0.61$. 
Increasing $\Gamma$ again to $5.0$ demonstrates the reversibility i.e. no hysteresis.
We refer to the end points of the reversible branch as RAMP$_{\Gamma=0.25}$ and RAMP$_{\Gamma=5}$ in our tomograhic analysis.

The third preparation protocol VIB is inspired by epitaxial growth: particles are deposited at a rate of approx. 15/second, 
while continuous sine vibration at $100\,\mathrm{Hz}$ and $\Gamma=5$ is applied; 
a similar method has been employed to create dense packings of ellipsoids \cite{donev:03}. 
Protocol VIB creates the densest packings in our study with $\phi=0.622$. 
Apparently the slow deposition and the continuous agitation 
of the surface layer enables the particles to explore more effectively the configurational phase space and 
thus find configurations with lower potential energy. 

Our range of mechanically stable packings of $\phi = 0.469$ to $0.622$  is in accordance with the range of $\phi=0.48$ to $0.64$ found
previously for tetrahedral dice and ceramic tetrahedra particles prepared by shaking and a water-fluidized bed  \cite{baker:10}. 
Our {\it range of volume fractions} is however significantly lower than the 
{\it single values} of $\phi = 0.76 \pm 0.02$ \cite{jaoshvili:10} and 
$\phi = 0.749 \pm 0.004$  \cite{zhao:12} reported for packings of tetrahedral dice.
It has been  found that weakly truncated corners do not increase maximal packing
fraction \cite{damasceno:12}, and rounded corners even decrease the packing fraction \cite{zhao:12}.
As the dice have more rounded corners than our particles,
the most likely explanation for the higher $\phi$ in \cite{jaoshvili:10,zhao:12} are different values of friction.

{\it Tomography analysis. --}
Tomographic reconstruction of the packings allows us to investigate the geometrical properties of the packing by recording
the positions and orientations of all tetrahedra. 
The numerical analysis of the volume data is based on an iterative  two-step algorithm: 
first, the approximate positions of the particle centroids are 
computed through the local maxima of cross-correlation with an inscribed sphere with inradius $R_i=a/(2\sqrt{6})$. Second, the 
orientation and exact position of each particle are determined using a steepest ascent gradient search. 
Detected particles are removed from the volume and the process is repeated with evenly gridded starting positions, 
until no more matches are possible. 
This approach succeeds in detecting more than 99.8\% of all tetrahedra  
(final reconstruction illustrated in Fig.~\ref{fig:particles} right).
More details can be found in the supplementary materials. 
The further analysis is restricted to an inner cylindrical region with diameter $11\,a$ and height $11\,a$, 
containing 4062 to 5342 particles (depending on $\phi$). The vertical variation of the local packing fraction 
in this core region is smaller than 0.002. The particle coordinates of all 22 experiments reported here can 
be downloaded from the Dryad repository \cite{note3}.

\begin{figure}
  \hspace{-0.4cm}
  \subfloat{\label{fig:gofr}\includegraphics[width=0.25\textwidth]{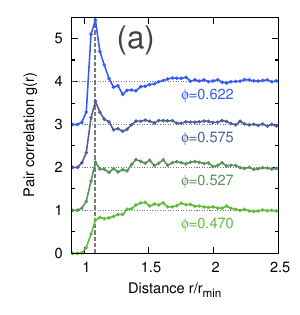}}
  \hspace{-0.4cm}
  \subfloat{\label{fig:fofr}\includegraphics[width=0.25\textwidth]{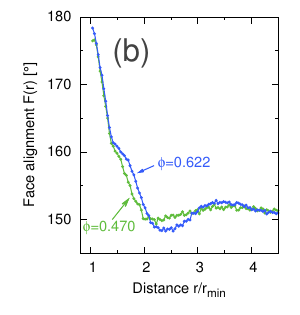}}
  \caption{(a) Pair correlation function g(r) for different packing fractions.
Offsets have been added for improved visibility.
 (b) Angular face correlation F(r).
(Color online)}
\label{fig:gofr_gofrpeaks}
\end{figure}

{\it Structural correlations. --}
A structural measure used both in jamming and in liquid theory is the pair correlation function $g(r)$, which
counts the number of neighbors at a given distance $r$, normalized by the binning volume and the average density. 
\refb{fig:gofr} shows $g(r)$ for a range of $\phi$.
The first peak is located at $r=0.44a$  (dashed line) 
and grows monotonely towards denser packings, indicating an increase of face-to-face contacts. 

The peak location is close but not identical to the minimal possible distance of $2 R_i = r_{min} \approx 0.408a$. 
Unlike the $g(r)$ of sphere packings, the left shoulder of the first peak 
has intrinsically a finite slope. This feature is a consequence of the lower probability of perfect face-to-face alignment 
compared to either slightly shifted face-to-face or low angle face-to-edge contacts. 
One consequence of this specific shape of the peak is that {\it compressed} packings of soft tetrahedra will not show the 
same square root scaling of contact number with volume fraction as observed in sphere packings \cite{vhecke:10}.   

$g(r)$ decays quickly after the first peak, showing no long-range translational ordering. This is in contrast to the $g(r)$ 
obtained from Monte Carlo (MC) simulations of tetrahedra\cite{haji-akbari:09,jiao:11} which revealed long-range structure 
at densities as low as $\phi=0.46$. The difference can be explained by the preparation: the MC simulations 
prepare packings without gravity and friction by slow compression of a liquid phase.
In contrast our frictional samples start to jam due to gravity at $\phi \approx 0.47$, 
after which geometric frustration and pressure in the pile restrict further alignment.

The orientational order can be quantified with an angular correlation function $F(r)$ (depicted in \refb{fig:fofr}), 
which measures the average relative orientation of face normals\cite{jaoshvili:10}.
$F(r)$ is computed by averaging $F_{ql}$  for all pairs of tetrahedra $q$ and $l$ whose centers have the distance $r$.
$F_{ql}$ is defined as the minimum of the pairwise dot product of the $4$ face normals $\vec{n}_{q}$ and $\vec{n}_{l}$:\\
$F_{ql} = \mathrm{min}(\vec{n}_{q,1 \dots 4} \cdot \vec{n}_{l, 1 \dots 4})$.
$F(r)$ reaches $180^\circ$ for a perfect face-to-face configuration and $150^\circ$ for a random configuration.
The angular correlations extend further which is in good accordance with 
previous experiments\cite{jaoshvili:10} and numerical results\cite{smith:10}. 
The development of a shoulder for the densest packing, starting at $r/r_{min}\approx 1.6$ may indicate an increase of 
trimers (3 tetrahedra aligned face-to-face), having a ideal centroid distance of $1.63\,r/r_{min}$.

{\it Contact numbers. --}
In order to study the origin of mechanical stability, we determine
the average contact number $Z$ of the tetrahedra.
As we cannot distinguish real contacts from close neighbors based on local force measurements \cite{majmudar:07}, 
we determine $Z$ from a model based fitting method introduced by Aste {\it et al.} \cite{aste:05} for sphere packings.
The supplementary material provides more details on this approach which does account for the 
finite resolution of tomography and detection.

\begin{figure}[t]
    \includegraphics[width=0.5\textwidth]{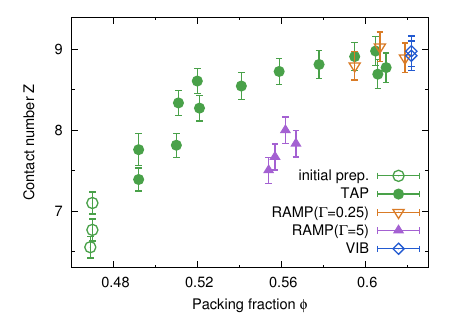}
    \caption{\label{fig:zofphi}Average contact number $Z$ as a function of packing fraction $\phi$ for different preparation protocols.
 (Color online)
}
\end{figure}

The resulting contact numbers at different densities and for the different preparation protocols are shown in 
\refb{fig:zofphi}. The packings prepared by TAP show $Z$ increasing monotonely with $\phi$. 
The contact numbers of the two densest packings, RAMP$_{\Gamma=0.25}$ and VIB, remain constant compared 
to the densest TAP sample. However, the packings prepared by RAMP$_{\Gamma=5}$ deviate distinctly: 
For comparable $\phi$, contact numbers are approx.~1.0 lower, confirming the influence of the preparation protocol. 
The initial loose packings show variations in $Z$ of the order 0.5.

{\it Constraints. --}
Contrary to spheres, $Z$ is not sufficient to evaluate the distance to isostaticity, 
because the number of mechanical constraints fixed by a contact depends on the specific contact geometry. 
Tetrahedra have 4 different types of contacts:  face-to-face (F2F) contacts, 
which are mechanically equivalent to 3 individual  point contacts, 
edge-to-face (E2F) contacts (equivalent to two point contacts) and the vertex-to-face (V2F) 
and edge-to-edge (E2E) contacts which correspond 
to a single point contact. 
All contacts impose 3 translational constraints, E2F contacts add 2 rotational constraints, F2F contacts probibit 3 different roationss.
As these constraints  are shared between two tetrahedra, we obtain the  
constraint numbers $C_{F2F}=3.0$, $C_{E2F}=2.5$, and $C_{V2F}=C_{E2E}=1.5$ (for more details please see supplements).
Multiplying these numbers with the according type-specific contact numbers gives 
the number of constraints per particle $C$. For an isostatic packing $C$ needs to be equal to the degrees of freedom each particle possesses, in the
case of tetrahedra this number is 6 \cite{note2}.

\begin{figure}[h]
  \includegraphics[width=0.5\textwidth]{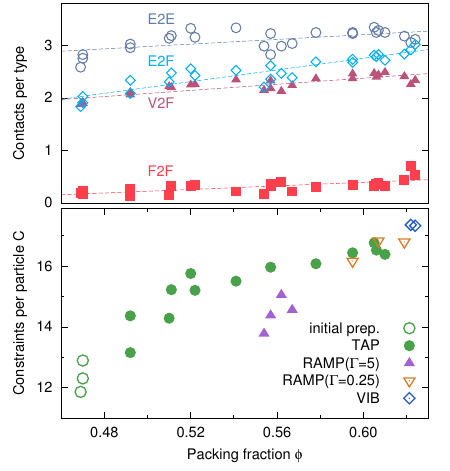}
  \caption{\label{fig:contacts_constraints} Top: Number of different contacts per particle as a function of packing fraction. 
The number of edge-to-face (E2F) contacts increases stronger than the other three contact types.
Bottom: Variation of the number of constraints per particle with packing fraction. The isostatic
limit corresponds to 6 constraints per particle.
(Color online)
}
\end{figure}

The contact type are determined by using the adjacent face-to-face-angle $\alpha_{FF}$ of each pair of 
contacting tetrahedra.
We e.g.~classify a contact as F2F if $\alpha_{FF}$  is in the range $176.9^{\circ}$ to $180^{\circ}$. 
The lower threshold of this range is based on an experimental calibration: 
We prepare a sample with 90 single face-to-face contacts and measure the cumulative distribution of $\alpha_{FF}$ in this particular sample. 
This distribution is well described by an error function with a point of inflexion at $m=178.2^{\circ}$ and a standard deviation 
$\sigma = 1.3^{\circ}$, suggesting an optimal threshold at $m - \sigma$.
Please see the supplementary material for full data and methods.

Figure \ref{fig:contacts_constraints} shows that the number of all four types of contacts
increases with  $\phi$. More specifically, the growth of the E2F contact number contributes about half of the total increase 
in $Z$. Together, the number of constraints per tetrahedra is between 12 and 18 and therefore much higher than the isostatic limit of 6. 
While this result does not contradict the isostatic packings found in simulations of frictionless tetrahedra \cite{smith:10,smith:11,jiao:11},
there is a clear disagreement with the experiments using frictional dice reported in \cite{jaoshvili:10}. There, however, 
the authors use the contact-specific constraint numbers pertinent to frictionless particles. If the contacts are treated as frictional,
the result will also be a hyperstatic packing. See also the discussion in \cite{smith:11}.

Both, the fact that tetrahedra packings are hyperstatic and that $C$ is not only a function 
of $\phi$ but does also depend on the packing protocol, necessitate expansions of contemporary theoretical approaches  
to static granular media. E.g.~the statistical mechanics approach to sphere packings by Song {\it et al.}
assumes a single inverse relationship between the mean free volume and contact number, independent of preparation
details \cite{song:08}. Similarly, the jamming paradigm \cite{liu:10,vhecke:10} is based on proximity to the
isostatic point and a power law relationship between $Z$ and $\phi$.  

{\it Conclusion. --}
We have shown that in jammed packings of frictional tetrahedra both the contact number 
and the number of mechanical constraints per particle $C$ 
grow monotonely with the packing fraction and depend on the preparation history.   
Contrary to earlier results, all packings were strongly hyperstatic with $C$ between 12 and 18 while each particle 
has only 6 degrees of freedom.
Our experiments involve non-spherical particles and consider friction explicitly, thus providing experimental support for 
extending the jamming paradigm towards real world granular media.

We thank U. Krafft, F. Schaller, K. Thomas, and S. Utermann for technical 
assistance, and A. Jaoshvili, F. Schaller, and A. Witt for helpful discussions. 
Furthermore, we acknowledge generous support from BP Exploration
Operation Company Ltd. within the GeoMorph research program.


\begin{thebibliography}{1}
\bibitem[{not({\natexlab{a}})}]{note1}
\bibinfo{note}{These values are obtained by inspecting the cross section of a
  silicone rubber imprint of the particles with an optical microscope.}
\bibitem[{not({\natexlab{c}})}]{note2}
\bibinfo{note}{Our definition of $C$ differs from previous definitions
  \cite{jaoshvili:10,smith:11,jiao:11} by a factor of 2 as it gives the number
  of constraints per particle.}
\end{thebibliography}
\end{document}


\title{Supplementary material to ``Jammed frictional tetrahedra are hyperstatic''}
\date{\today}
\author{Max Neudecker, Stephan Ulrich, Stephan Herminghaus, Matthias Schr\"oter}
\maketitle
\tableofcontents

\section{Particle detection}

\subsection{Preprocessing}
\begin{figure}[!ht]
  \centering
    \includegraphics[width=.95\textwidth]{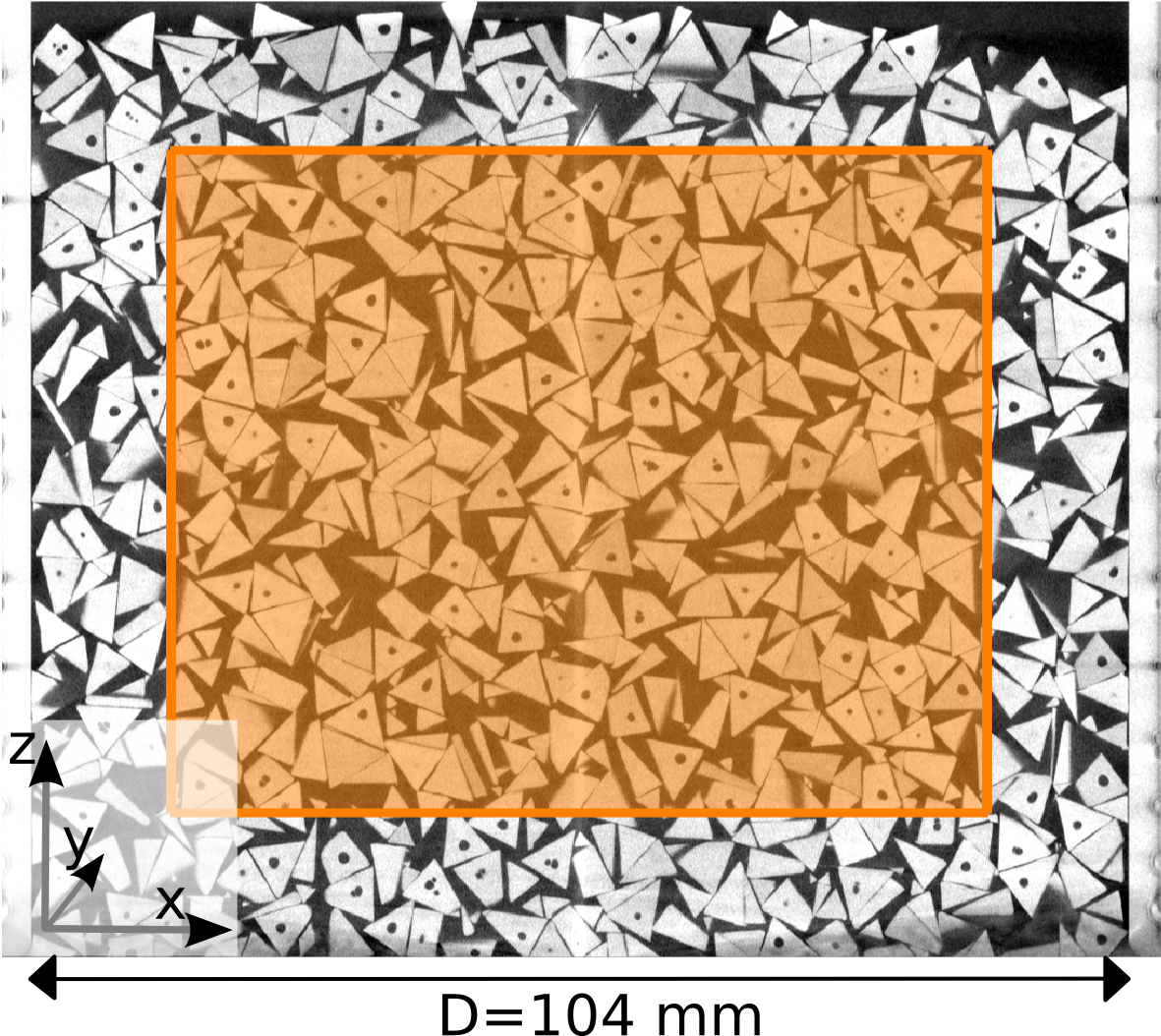}
    \caption{Exemplary volume slice at central x-z-plane ($y=53\,\textrm{mm}$). 
The inner region of interest (ROI) used for analysis is shown as an orange 
rectangle. The cylinder walls are visible, the container bottom is approx. $10\,\textrm{mm}$ below.}
    \label{fig:sliceroi}
\end{figure}

A typical reconstructed volume dataset consists of 1152x1152x1076 voxels with a 
spatial resolution of $100\,\mathrm{\mu m}$ per voxel. Figure~\ref{fig:sliceroi} shows a cut through the central x-z-plane, 
where the region of interest (ROI) is marked with a rectangle. Note that the particles outside this region are detected as well, 
but only particles inside the ROI are used in further analysis. 
The first step is a radially varying binarization using Otsu's threshold
\cite{otsu:79}, which accounts for the radial brightness decay typically observed in our tomograms.

The second step removes shrinkage cavities, which are inherent to
injection molding, as follows:
\\
\begin{enumerate}
\item Inverting the binary image
\item Identifying connected white regions using the MATLAB labeling algorithm
\texttt{bwlabeln} \cite{sedgewick:98}.
\item Removing regions with voxel volume $< 20000$ (hole size threshold)
\item Inverting the image again
\end{enumerate}

\subsection{Positions}

The approximate centroid positions of the tetrahedra with sidelength $a=7\,\mathrm{mm}$ are detected by cross-correlation of a sample sphere with the binarized volume. The radius of the sphere is set to the tetrahedra inradius $R_{i}=a/(2 \sqrt{6})$ which corresponds to $14$ integer voxels. Thresholding the resulting correlation volume with .98 of the maximum value results in a set of $N_c$ center regions, whose centroids $\vec{c}_i \,(i=1...N_c)$ are good estimates for the tetrahedra centroids.

\subsection{Orientation}
\begin{figure}[!ht]
  \centering
  \includegraphics[width=0.4\textwidth]{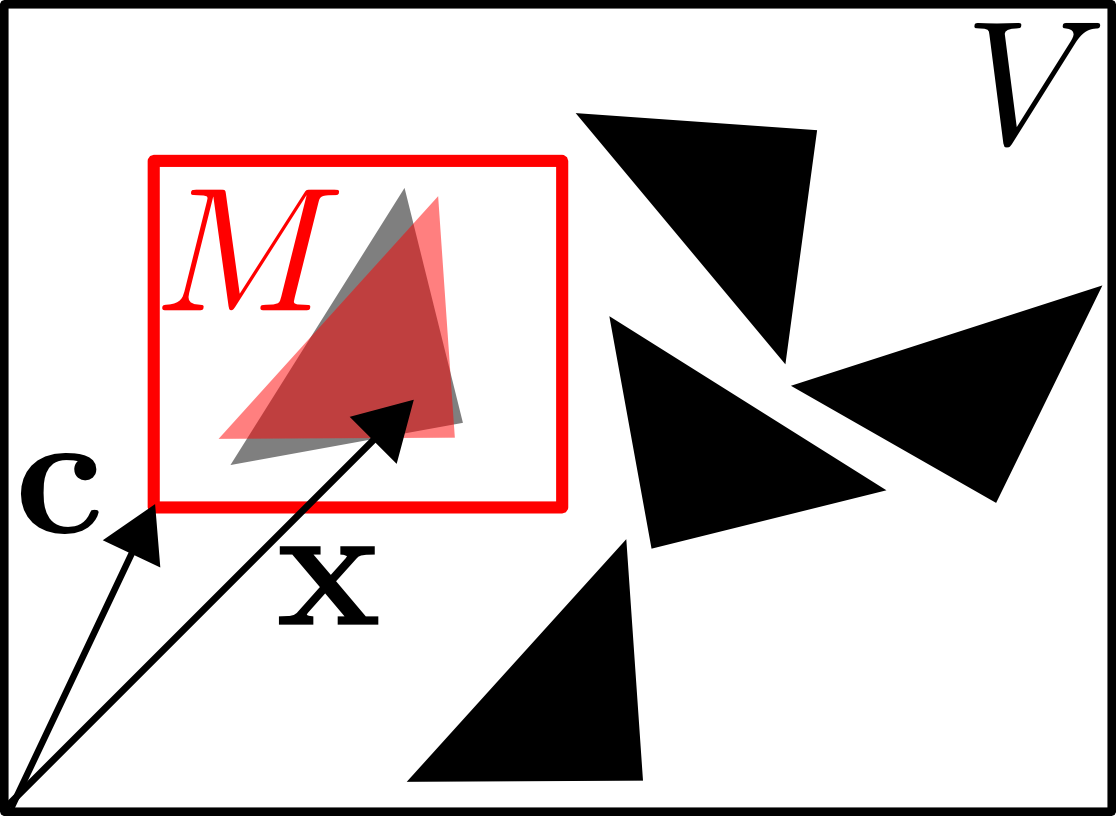}
  \caption{Sketch of model $M$ (red triangle) and volume $V$ (samples outside $M$ as black triangles, inside $M$ shaded grey): 
a voxel $\mathbf{x}$ given in coordinates of $V$ is shifted by $\mathbf{c}$ into the origin of $M$ 
before the convolution is applied. 
In this example, turning the model a few degrees counter-clockwise would increase the overlap of model and volume.
}
  \label{fig:convolution_sketch}
\end{figure}

To find the orientation of the individual tetrahedra, we use a binary model tetrahedron $\mathbf{M}$ 
which we grow, translate and rotate while maximizing the voxel overlap with the binary tomography data $\mathbf{V}$
(see Fig.~\ref{fig:convolution_sketch}).
The binary images $\mathbf{M}$ and $\mathbf{V}$ are defined as functions which map from $\mathbb{R}^3$ 
to the interval $[0,1]$. Note that while $\mathbf{V}$ is defined in the complete volume, $\mathbf{M}$ 
contains only one tetrahedron with a much smaller definition domain $D_m$. 

This allows to define the overlap as a spatial convolution\cite{soille:2005}, usually denoted by~$*$. The value 
of the convolution at a given voxel $\mathbf{x} \in \mathbf{V}$ and given offset $\mathbf{c}$ is computed by 
taking the sum of pixels of $\mathbf{V}$, weighted by $\mathbf{M}$ in the subset $D_m$, where $D_m$ is shifted 
by the offset $\mathbf{c}$:

\begin{equation}
  (\mathbf{V}*\mathbf{M})(\mathbf{c}) = \sum_{x=D_m} \mathbf{V}(\mathbf{x})\mathbf{M}(\mathbf{x-c})\,.
  \label{eq:convolution}
\end{equation}

\pagebreak

\begin{wrapfigure}[36]{r}{0.54\textwidth}
  \includegraphics[width=0.53\textwidth]{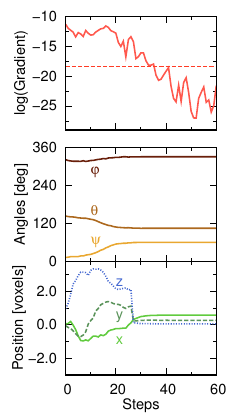}
  \caption{From top to bottom: Evolution of the gradient (solid line) and threshold for convergence (dotted line); orientation 
(Euler angles $\phi$, $\theta$ and $\psi$), and position ($x$,$y$,$z$) relative to the previously detected centroid until convergence.}
  \label{fig:posangevolution}
\end{wrapfigure}

We implemented a steepest-gradient search algorithm in MATLAB  to address the registration problem directly: The model $\mathbf{M}$ is 
placed at the approximate centroid position (x,y,z) and set to a random rotation 
(given as 3 Euler angles $\phi$, $\theta$ and $\psi$). 
The objective function (``Zielfunktion'') to be maximized is the magnitude of the convolution function as 
defined in Eq.~\ref{eq:convolution}.
In each algorithm step, the finite gradient approximation for all 6 parameters is computed, and a move 
in the gradient direction is performed. The algorithm starts out with a shrunk model (sidelength of $4.2\,\mathrm{mm}$ 
and grows the model to the assumed sidelength of $7\,\mathrm{mm}$, improving robustness and performance. 
The convergence criterium is the magnitude of the gradient, which is required to be lower than a pre-defined 
threshold for more 
than 20 successive steps (as shown in Fig.~\ref{fig:posangevolution}). 
When this exit condition is reached, and the overlap of model and volume is at least 96\% of the theoretical maximum, 
the particle is marked as detected and removed from the volume.

\begin{figure}[!ht]
\centering
      \subfloat[Step 1]{\includegraphics[width=0.32\textwidth]{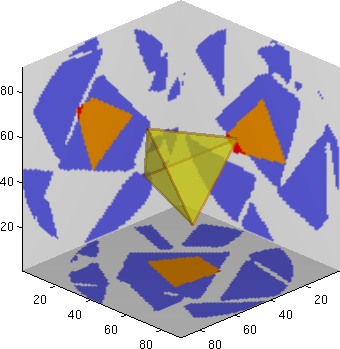}}
      \subfloat[Step 21]{\includegraphics[width=0.32\textwidth]{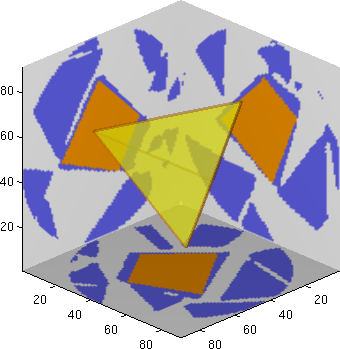}}
      \subfloat[Step 41]{\includegraphics[width=0.32\textwidth]{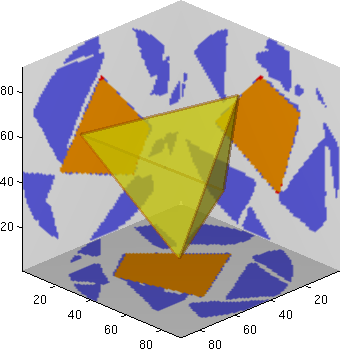}}
      \caption{Snapshots corresponding to the parameter evolution in Fig.~\ref{fig:posangevolution}. 
        The background depicts cuts through the 3 cartesian directions at the current (x,y,z)-position, 
        projected outwards for better visibility. Color coding: gray is background (empty space), 
        blue is tetrahedra-material, orange corresponds to model-tetrahedra-overlap, 
        and red to model-background-overlap. 
        The foreground shows the current position and orientation of the model (yellow). }
      {\label{fig:posangsnapshot}
    }
\end{figure}

 A typical evolution of the parameters during the detection of one tetrahedron is shown in 
Fig.~\ref{fig:posangevolution}, and Fig. \ref{fig:posangsnapshot}. Convergence is reached typically after 40-80 steps, 
depending on the initial parameters and the local neighborhood.

In some cases, particularly in dense packings with many face-to-face contacts, the particle detection does not converge 
in the first place. 
Therefore, the search may be repeated up to 10 times with evenly gridded starting positions and random angles, finally
 achieving a detection rate in the range of 99.80\%-99.95\% inside our ROI.

\pagebreak

\section{Contact analysis}

Inter-particle contacts cannot be determined directly from tomographic data because the result is affected 
by particle polydispersity, imaging artefacts and the finite accuracy of particle detection. 
However, Aste {\it et al.}~\cite{aste:05} introduced a method that can provide a physically justified average contact number 
for spheres of diameter $d$  (assuming that experimental errors are distributed gaussian).
It defines a number of neighboring particles $n(r)$ which includes all particles with a  
center to center distance smaller or equal to $r$. In the case $r<d$, $n(r)$ can be 
described by multipying the average contact number $Z$ 
with a cumulative normal distribution  $\Phi_{\mu,\sigma}(r)$: 
\begin{equation}
  \Phi_{\mu,\sigma}(r) = \frac{1}{ \sigma \sqrt{2 \pi}} \int_{-\infty}^r \exp\left(-\frac{(x-\mu)^2}{2\sigma^2} \right)\textrm{dx}
  \label{eq:cumgauss}
\end{equation}

The mean $\mu$ provides an estimate for the bead diameter $d$, and
the variance $\sigma$ corresponds corresponds to the combined effects 
of polydispersity and uncertainty of tomography and particle detection.  

\begin{figure}[!ht]
  \centering
  \includegraphics[width=0.7\textwidth]{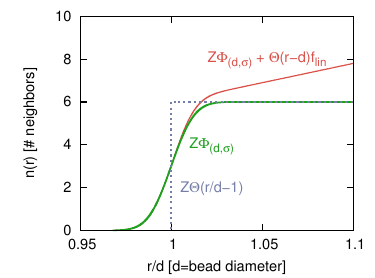}
  \caption{Model for the number of apparent neighbors $n$ as a function of the center-to-center-distance $r$ in a packing 
of spheres with diameter $d$ and contact number $Z=6$. The resulting contact number can be read off after deconvoluting 
the function by the scaled cumulative normal function $\Phi_{d,\sigma}$.}
    \label{fig:zofa_spheres}
\end{figure}

For $r>d$ there is an additional term to  $Z \Phi_{\mu,\sigma}(r)$ which describes
the growing contribution from ``spurious'' contacts: near neighbors which are close but not in contact. 
The latter part can be approximated by a linear function $f_{lin}(r) = m\cdot(r-d)$ with slope $m$ combined 
with the Heaviside step function $\Theta(r-d)$, leading to the complete model for $n(r)$:
\begin{equation}
  n(r) = Z \cdot \Phi_{d,\sigma} + \Theta(r-d) \cdot f_{lin}
  \label{eq:nrmodel}
\end{equation}
Figure \ref{fig:zofa_spheres} shows an illustration of the model for spheres with a contact number $Z=6$.

We transfer the model to tetrahedra packings as follows: Instead of counting neighbors within a distance $r$, particles 
are scaled by a ``virtual'' side length $a_v$ and the number of intersections $n(a_v)$ is 
counted\footnote{A substantial speedup of the analysis is reached by using neighbor cell lists and testing only particles 
within a diameter of the circumsphere.}, as shown in Eq.~\ref{eq:nav}:
\begin{equation}
  n(a_v) = Z \Phi_{a_{\mu},\sigma} + \Theta(a_v-a_{\mu}) \cdot m \cdot (a_v-a_{\mu})
  \label{eq:nav}
\end{equation}
with the estimated sidelength $a_{\mu}$.

\begin{figure}[!ht]
  \centering
  \subfloat[$\phi=0.470 \,, Z=7.1$]{\label{fig:zofa_taps0g4}\includegraphics[width=0.5\textwidth]{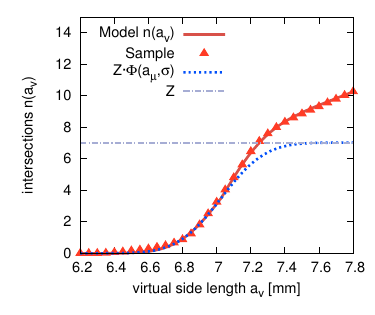}}           
  \subfloat[$\phi=0.600 \,, Z=8.8$]{\label{fig:zofa_run1down}\includegraphics[width=0.5\textwidth]{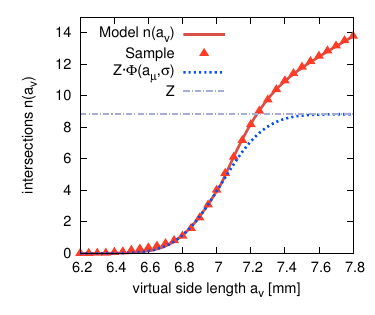}}           
  \caption{Contact number analysis of tetrahedra packings for (a) loose and (b) dense packing. The data from the samples 
(red triangles) is well approximated by the model $n(a_v)$ (solid red line) using only $Z$ and $m$ as fit parameters.
Deconvolution into the cumulative gauss $\Phi_{a_\mu,\sigma}$ (dotted blue line) 
and the contribution $f_{lin}$ allows to read off the contact number $Z$ (dash-dotted grey line).
}
  \label{fig:zofa_tetrahedra}
\end{figure}

Exemplary contact curves for loose and dense packings are presented in Fig.~\ref{fig:zofa_tetrahedra}. 
Firstly, the side length $a_{\mu}$, the contact number $Z$, the variance $\sigma$ and the linear slope $m$ are fitted to the data, 
using the \texttt{gnuplot} implementation of the non-linear least-squares Levenberg-Marquardt algorithm. 
Since particles and experimental setup are identical for all 25 samples, we determine the average sidelength 
$\bar{a}_{\mu}=(7.02 \pm 0.03)\,\mathrm{mm}$ and the average variance $\bar{\sigma}=(0.206 \pm 0.015)\,\mathrm{mm}$ from 
all fits ( $\pm$ errors are standard deviations). In a second step, these parameters are fixed, leaving $Z$ and $m$ 
as the only free parameters.

\subsection{Estimation of error bars}

The error estimate $\Delta Z$ of the contact number $Z$ can be calculated by the propagation of error principle: 
$Z$ is a function of the estimated sidelength $\bar{a}_{\mu}$ and the distribution width $\bar{\sigma}$, with the variances $\Delta \bar{a}_{\mu}$ and $\Delta \bar{\sigma}$. 
We denote the errors of the mean values with $\tilde{a}_\mu$ and $\tilde{\sigma}$. 
The error of the mean value, $\tilde{a}_\mu$, is computed by $t_s\cdot \frac{\Delta \bar{a}_\mu}{\sqrt n}$ (with the Student-t-distribution factor $t_s=2.13$ 
according to $n$=20 samples and 95\% confidence), respectively for $\tilde{\sigma}$.

The error of $Z$ is then computed as follows:

\begin{equation}
  \Delta Z = \sqrt{\left(\frac{\partial Z}{\partial \bar{a}_{\mu}}\cdot \tilde{a}_\mu \right)^2 + \left(\frac{\partial Z}{\partial \bar{\sigma}}\cdot \tilde{\sigma}\right)^2 }
\end{equation}

where the partial derivates are approximated by the difference quotient approximation

\begin{equation}
  \frac{\partial Z}{\partial \bar{\sigma}} \approx \frac{Z(\bar{\sigma}) - Z(\bar{\sigma}+\Delta \bar{\sigma})}{\Delta \bar{\sigma}} \,.
\end{equation}
 (respectively for $\bar{\sigma}_{\mu}$).

\section{From contacts to constraints}

There are 4 different types of contacts in tetrahedra packings, as table \ref{tab:contact_types} illustrates:  
face-to-face (F2F) contacts, edge-to-face (E2F) contacts and the point contact configurations vertex-to-face 
(V2F) and edge-to-edge (E2E). Vertex-Vertex or Vertex-Edge contacts are not observed in practice.

The number of constraints fixed at a specific type of contact is best evaluated by a thought experiment
where one of the two particles is kept fixed. In the presence of friction, all contact types will then block three 
translational degrees of the second tetrahedra: one normal and two tangential. The contact types do however differ
in the amount of blocked roational degrees. A frictional face-to-face contacts blocks 3 rotations: one around the surface normal (by friction)
and 2 rotations around the two axis standing perpendicular on the surface normal (by non-overlap). An E2F contact blocks only one rotation
perpendicular to the surface normal. Finally the pointlike contacts V2F and E2E don't block any rotation at all. 
As the constraints are shared between two tetrahedra, we obtain the  
constraint multipliers $C_{F2F}=3.0$, $C_{E2F}=2.5$, and $C_{V2F}=C_{E2E}=1.5$. 
Multiplying these numbers with the according type-specific contact numbers in a specific configuration
gives the number of constraints per particle $C$.

\begin{table}[!ht]
  \begin{tabular}{lllll}
    \hline
    \textbf{Type} & face-to-face & edge-to-face & vertex-to-face & edge-to-edge \\
    \hline
    \textbf{Example} &
    \includegraphics[width=0.17\textwidth]{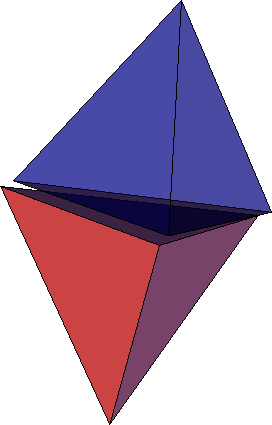} & 
    \includegraphics[width=0.17\textwidth]{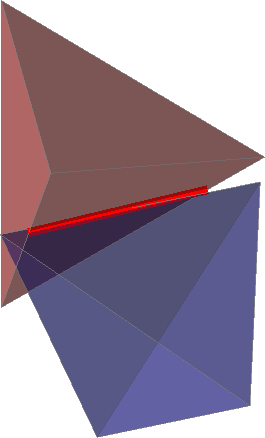} & 
    \includegraphics[width=0.17\textwidth]{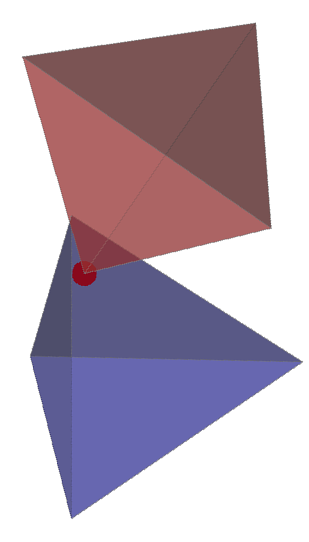} & 
    \includegraphics[width=0.17\textwidth]{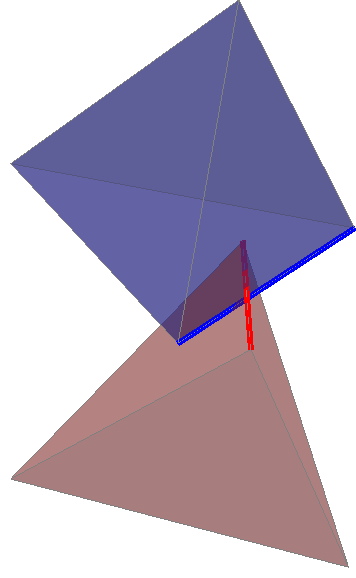} \\
    \hline
    \textbf{Constrained}\\\textbf{DOF} &
     3 trans. \& 3 rot. &
     3 trans. \& 2 rot. &
    3 translational &
    3 translational \\
    \hline
    \textbf{Constraints}\\\textbf{per Particle} &
    3.0 & 
    2.5 &
    1.5 &
    1.5 \\
    \hline
  \end{tabular}
  \caption{From top to bottom: Contact geometries as classified from a experimental tetrahedra packing; constrained DOF; 
constraint multipliers $C_{F2F}$, $C_{E2F}$, $C_{V2F}$, $C_{E2E}$;}
  \label{tab:contact_types}
\end{table}

\subsection{Analysis of the contact geometry}

\begin{figure}[!ht]
  \centering
    \includegraphics[width=0.5\textwidth]{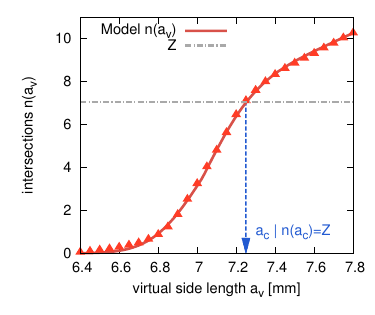}
    \caption{The side length $a_c$ with $n(a_c)=Z$ is given by the intersection of the model $n(a_v)$ with $Z$.}
    \label{fig:zafitinverse}
\end{figure}

In order to determine local contact geometries, all tetrahedra of one sample are scaled to the ``contacting'' side length $a_c$, which makes the corresponding $n(a_c)=Z$ consistent (see Fig.~\ref{fig:zafitinverse}). Then the pairwise contact geometry of all tetrahedra with their intersecting neighbors is analysed.

Our classification algorithm starts by checking the face-to-face angle $\alpha_{FF}$, defined as 
$\cos^{-1}(min(\mathbf{n}_{i}^q \cdot \mathbf{n}_{j}^k))$ for two contacting tetrahedra q,k, 
and all face pairs $i,j \in [1..4]$ 
(see also \cite{jaoshvili:10,smith:11}). Visually speaking, this is the angle between normals 
of adjacent faces, which is 180\textdegree~for 
perfect alignment. Because of our finite resolution, we classify  
a contact as F2F if $\alpha_{FF}$ is higher than a certain threshold value $\alpha_{FF}^{min}$.

\subsubsection{Threshold choice for F2F and E2F contacts}
\begin{figure}[htbp]
    \subfloat[Tomographic reconstruction of a F2F sample]{\label{fig:tetontetsample}\includegraphics[width=0.48\textwidth]{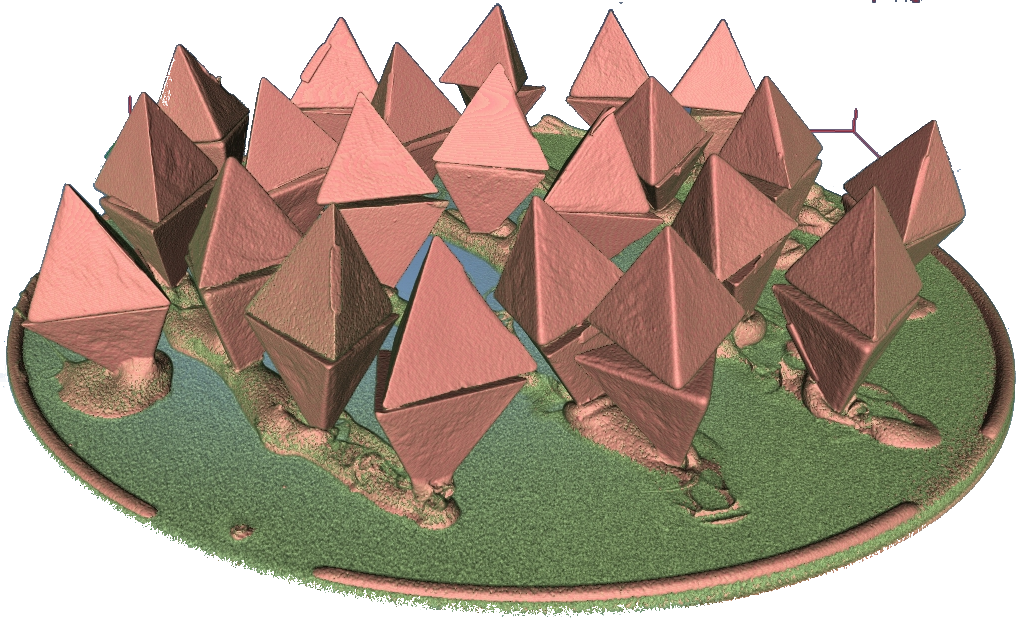}}           
    \hspace{0.5cm}
    \subfloat[Cumulative distribution of the number of face-to-face contacts per particle. 
    The chosen threshold is marked by dotted grey line]{\label{fig:fanglehist}\includegraphics[width=0.48\textwidth]{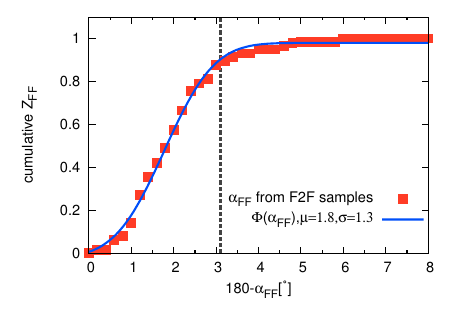}}        
    \caption{Sample containing only F2F contacts for determination of threshold $\alpha_{FF}^{min}$.}
\end{figure}

It has been shown\cite{smith:11} that an arbitrary choice of $\alpha_{FF}^{min}$ can lead to a physically 
infeasible constraint number\cite{jaoshvili:10}, therefore the threshold must be chosen carefully.
To this aim, five different samples containing only face-to-face contacts (``F2F sample'') are prepared by 
glueing one tetrahedron corner-down to a plate and adding another tetrahedron face-down on the top face of the first. 
A tomographic reconstructions of one of the samples is shown in Fig.~\ref{fig:tetontetsample} and the cumulative 
distribution of $\alpha_{FF}$ from all samples (containing 90 tetrahedra pairs) is given in Fig.~\ref{fig:fanglehist}. 
We find that a cumulative normal function as in Eq.\ref{eq:cumgauss} with mean $\mu=1.8$\textdegree~and 
variance $\sigma=1.3$\textdegree~is a good model for the distribution of the face-to-face angle $\alpha_{FF}$.

We therefore chose the threshold $\alpha_{FF}^{min}=3.1$\textdegree. 
This also applies for the classification of an edge-face contact, that is, the angle enclosed between an 
edge and a face in contact must be smaller than $\alpha_{EF}=\alpha_{FF}^{min}$.
Vertex-to-face contacts can be identified by analysing if one vertex of tetrahedron $A$ is inside the 
contacting tetrahedron $B$ (these intersections are possible due to the scaling to $a_c$). 
Any remaining contacts can then be ascribed to edge-to-edge contacts.

\bibliography{supplement}
\bibliographystyle{plain}